\documentclass[twocolumn,runningheads,natbib]{svjour2}
\smartqed  

\usepackage{epsf}
\usepackage{bm}
\usepackage{graphicx}

\newcommand{\la}{\;\raise0.3ex\hbox{$<$\kern-0.75em\raise-1.1ex\hbox{$\sim$}}\;}
\newcommand{\ga}{\;\raise0.3ex\hbox{$>$\kern-0.75em\raise-1.1ex\hbox{$\sim$}}\;}
\newcommand{\sla}{\;\raise0.55ex\hbox{\scriptsize$<$\kern-0.75em\raise-1.1ex\hbox{$\sim$}}\;}
\newcommand{\sga}{\;\raise0.55ex\hbox{\scriptsize$>$\kern-0.75em\raise-1.1ex\hbox{$\sim$}}\;}
\newcommand{\ssim}{\;\raise0.3ex\hbox{\tiny$\sim$}\,}
\newcommand{\sapprox}{\;\raise0.3ex\hbox{\tiny$\approx$}\,}

\journalname{Astrophysics and Space Science}

\begin{document}

\title{Nucleon superfluidity versus thermal states of \\
      isolated and transiently accreting neutron stars
\thanks{ This work is supported by Polish MEiN (grant no. 1P03D-008-27),
         by Russian Foundation for Basic Research (grants 05-02-16245, 05-02-22003),
         by the Russian Federal Agency for Science and Innovations (grant NSh 9879.2006.2),
     and by Russian Science Support Foundation.
       }
}

\titlerunning{Nucleon superfluidity vs thermal emission of SXTs}

\author{K.P.Levenfish and P.Haensel}

\authorrunning{Levenfish, Haensel}

\institute{ K.P.Levenfish \at Ioffe Physical Technical Institute,
            St.-Petersburg, Russia \\
            \email{ksen@astro.ioffe.ru}
      \and
       P.Haensel \at N.Copernicus Astronomical Center, Warsaw, Poland \\
       \email{haensel@camk.edu.pl}
}

\date{Received: / Accepted: }

\maketitle

\begin{abstract}
The properties of superdense matter in neutron star (NS) cores
control NS thermal states by affecting the efficiency
of neutrino emission from NS interiors. To probe these
properties we confront the theory of thermal evolution of NSs with
observations of their thermal radiation. Our observational
basis includes cooling isolated NSs (INSs) and  NSs in
quiescent states of soft X-ray transients (SXTs). We find that the data
on SXTs support the conclusions obtained  from the analysis of
INSs:  strong 
proton superfluidity with  $T^{\rm max}_{cp} \ga 10^9$ K 
should be present, while
mild 
neutron superfluidity with $T^{\rm max}_{cn} \approx 2\times
(10^8$--$10^9)$ K is ruled out in the outer NS core.
Here $T^{\rm max}_{cn}$ and $T^{\rm max}_{cp}$ are the maximum
values of the density dependent critical temperatures of
neutrons and protons.
The data on SXTs suggest also that: {\sl(i)}\
cooling of massive NSs is enhanced by neutrino emission more
powerful than the emission due to Cooper pairing of neutrons;
{\sl(ii)}\ mild neutron superfluidity, if available,
might be present only in inner cores of massive NSs.
In the latter case SXTs would exhibit
dichotomy,
i.e. very similar SXTs
may evolve to very different thermal states.
\end{abstract}

\maketitle

\section{Introduction}
\label{Sec: Introduction}

Neutron stars are very compact; their cores contain matter
with density $\rho$ a few times larger than
the standard nuclear matter density $\rho_0=2.8 \times 10^{14}$
g~cm$^{-3}$. Many properties of this matter cannot be
calculated precisely or studied in laboratory experiments.
However, these properties can be constrained by comparing
neutron star theory with observations; see e.g., \cite{yp04} and
\cite{pgw06}, for recent reviews.

In this paper, we describe current contraints on composition
and superfluidity of neutron star cores, which can be obtained
by comparing calculated thermal states of neutron stars with
observations of thermal radiation from middle-aged INSs and
NSs in SXTs in quiescent states. INSs are thought to cool
gradually from initial hot states via neutrino emission from
NS interiors  (at age $\la 10^5$ years) and via surface photon
emission (at the elder age; the so called neutrino and photon
cooling stages, respectively). NSs in SXTs will be assumed to
support their warm states owing to deep crustal heating
(pycnonuclear reactions) in accreted matter (see
\citealt{brown98}). Their thermal energy  is partly emitted by
neutrinos and partly by the surface radiation. In both cases,
thermal states of NSs are very sensitive to composition and
superfluidty of matter in their cores. Although INSs and SXTs
are different objects, their observations allow one to test
the same physics of superdense matter (Yakovlev et al.\ 2003;
2004).

{\it Composition}
and {\it superfluidity of baryons} in
NS cores are main regulators of NS thermal states.
These properties are thought to be the same for a given density
in
all NS cores.
The composition determines
dominant processes of neutrino cooling
in NSs.
Superfluidity, if it appears, reduces
rates of neutrino pro\-ces\-ses at work,
but opens an additional
neutrino mechanism
associated with Cooper pairing of ba\-ry\-ons.
Superfluidity  affects also the NS heat capacity.
Other regulators of
NSs thermal states (composition of NS heat-blanketing envelopes,
strength and geometry of NS magnetic field, etc.)
are also important but, taken alone, do not allow one
to reconcile theory of NS thermal state with observations
(see, e.g., \citealt{yp04}).
Moreover, they may vary from one star to another.
We will neglect them, for clarity.

\begin{figure*}[th]
\begin{picture}(210,80)(30,30)
\put(29,25){\includegraphics[width = 17cm]{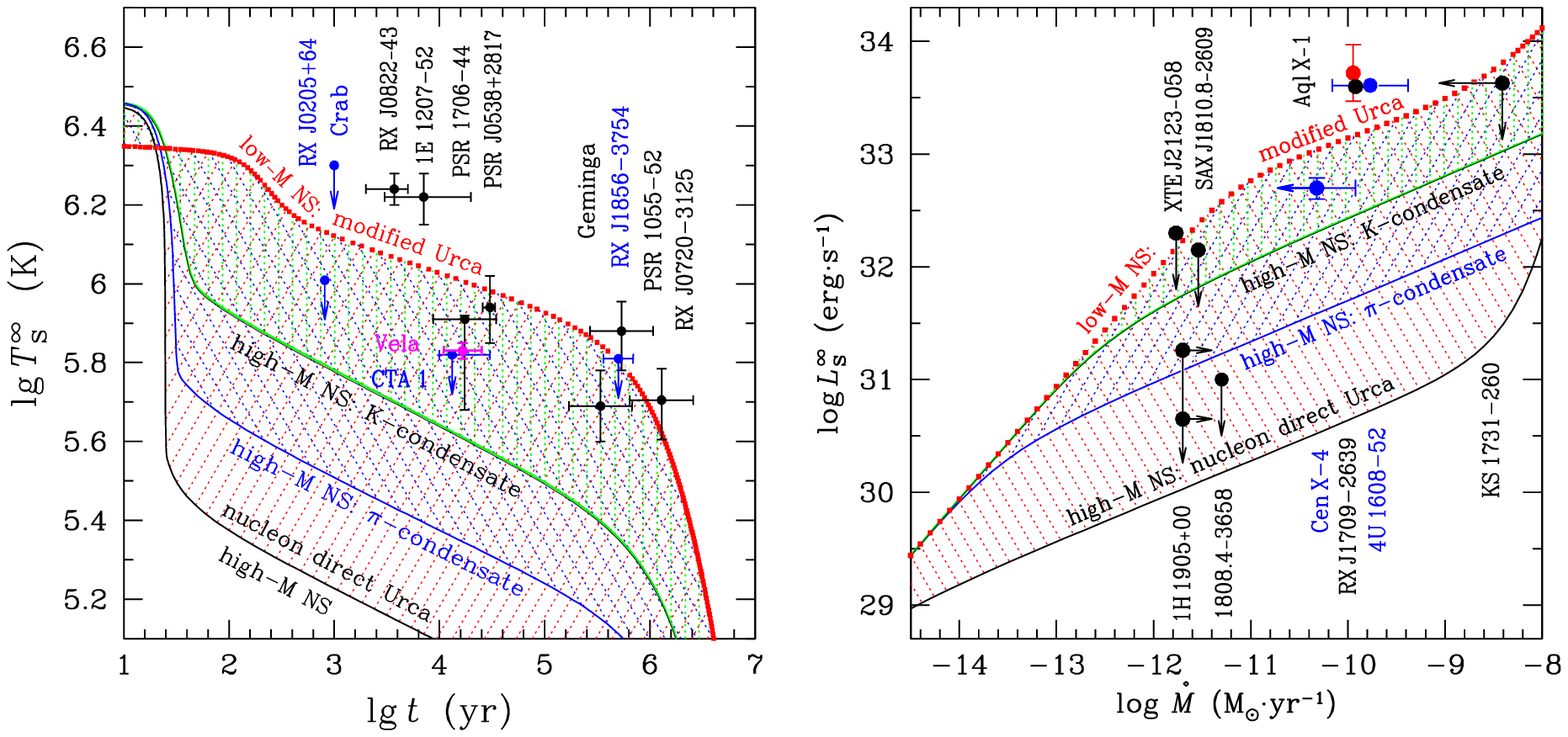}}
\end{picture}
\caption{
    Thermal states of nonsuperfluid NSs compared to observations.
    Left: effective surface temperatures of INSs, redshifted
    for a distant observer, versus their age.
    Right:
    redshifted photon luminosities of NSs in SXTs in quiescence
    versus time-averaged mass accretion rate.
    The dotted curves refer to the basic NS model (a
    non-superfluid low-mass NS which cools slowly through
    the modified Urca process).
    Three solid curves on each panel display scenarios with the enhanced
    neutrino cooling
    (maximum-mass NSs
    with inner cores containing --
    from  top  to  bottom --
    kaon condensates, pion condensates, and
    nucleons with large proton fraction, sufficient to open direct Urca
    process).  Hatched regions between the basic curve and any
    solid curve can be filled by curves of NSs
    with different masses, from $\sim 1\,M_\odot$ to
    the maximum one, for a corresponding NS composition.
}
\label{Fig:exot}
\end{figure*} 

\begin{figure*}[th]
\begin{picture}(210,80)(30,30)
\put(29,25){\includegraphics[width=17cm]{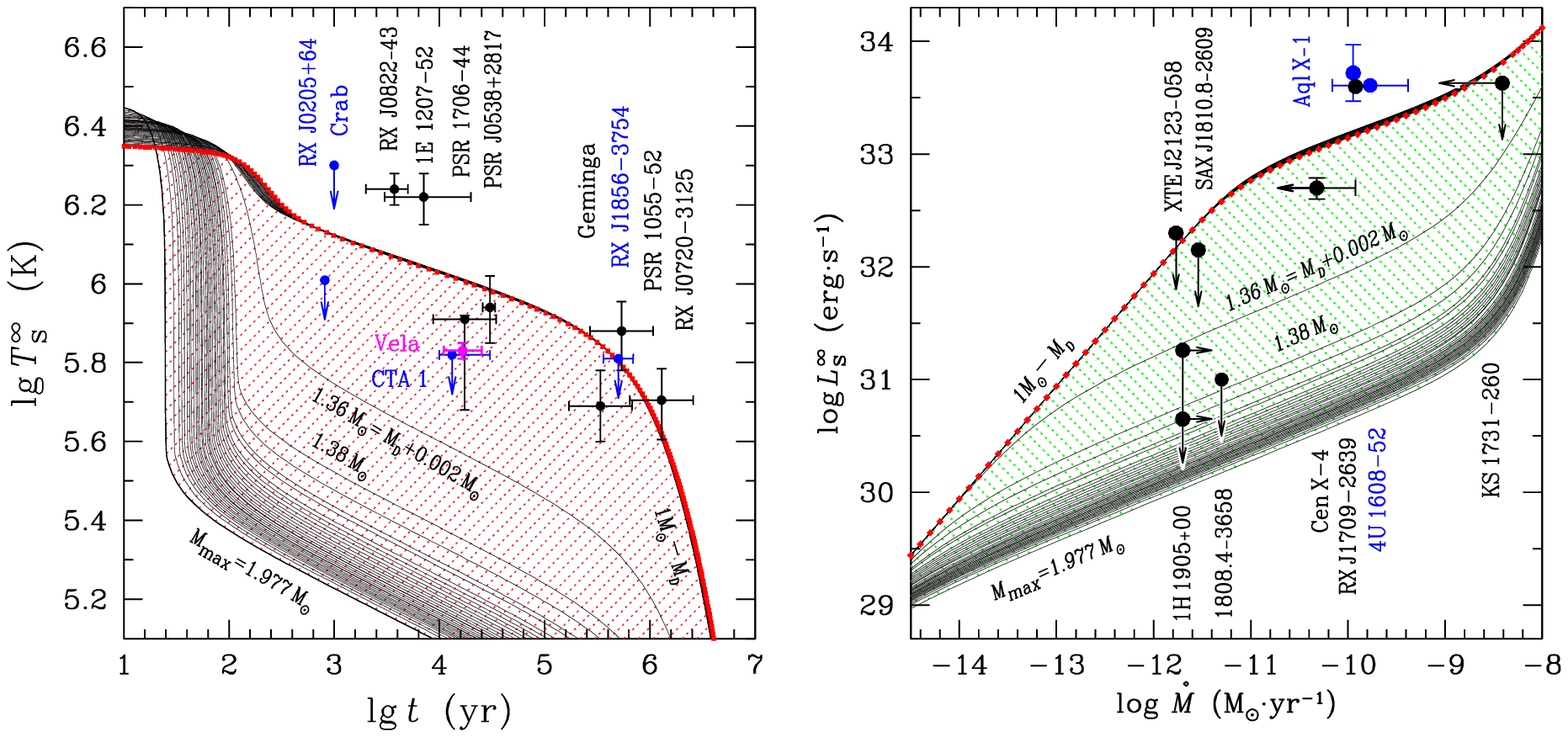}}
\end{picture}
\caption{Thermal states of nonsuperfluid INSs and SXTs with nucleon cores
based on an EOS of Prakash et al.\ (1988).
This EOS opens the nucleon direct
Urca process at $\rho \ge \rho_{\rm D} = 2.8\rho_0$, i.e., in inner cores of
NSs with $M \geq M_{\rm D}=1.358\, M_\odot$.
The thick dotted basic curves
are the same as in Fig.~\protect{\ref{Fig:exot}}.
From  top to bottom,  thin solid curves correspond to
NSs with masses growing from
$1\,M_\odot$ to $M_{\rm max}=1.977\,M_\odot$, with a step of $0.02\,M_\odot$.
An upper dense bundle of curves in each panel contains low-mass stars with
$M<M_{\rm D}$, while a less dense bottom bundle contains
NSs with $M > 1.06 \, M_{\rm D}$.
}
\label{Fig:none}
\end{figure*} 

\section{ Composition of NS cores}
\label{Sec: Composition}

An NS core can be divided into the outer core ($\rho \la $ a few$\,\rho_0$)
and the inner one (higher $\rho$).
The outer core consists mostly of neutrons,
with a small (a few \%) admix\-tu\-re of protons and
leptons. Low-mass NSs have only the outer core.
In the absence of nucleon superfluidity, they
cool down via the modified Urca processes and weaker neutrino
processes of nucleon-nucleon bremsstrahlung.

More massive NSs possess also the inner core whose
composition is largely unknown. According to different hypotheses,
the inner core may contain nucleons, hyperons, pion or kaon condensates,
or quarks. Even if composed of
nucleons and leptons, superdense matter can have a large
fraction of protons. In each of these
cases, neutrino emission
processes of direct Urca type become allowed,
much more powerful than the basic  modified Urca processes.
The most powerful is the nucleon (or hyperon) direct Urca
process. Neutrino reactions
with pions (kaons) are about two
(four) orders of magnitude weaker.

In the presented figures we compare
various theoretical predictions of
thermal states of INSs and NSs in SXTs with
observations. For INSs, we plot the redshifted effective
surface temperature $T_{\rm S}^\infty$ versus NS age $t$.
For SXTs, we show the redshifted surface luminosity
$L_{\rm S}^\infty$ of NSs in quiescence
versus time-averaged mass accretion rate $\dot{M}$.
We assume that NSs in SXTs are in thermal steady-states,
with the deep crustal heating (for a corresponding $\dot{M}$)
balanced by neutrino and photon emission; see \cite{ylh03} for details.
The deep crustal heating is calculated using the
model of \cite{hz90}.
In the majority of cases NS interiors are nearly isothermal
owing to high thermal conductivity, with the main temperature
gradients located in the heat-blanketing envelope near the NS surface.
We omit technical details because of space restrictions.
Observations of INSs are the same as in \cite{kyg02}.
The SXTs Aql X-1, 4U 1608-52, SAX 1808.3-3658 and Cen X-4
are described in \cite{ylh03}. The data on
XTE J2123-058, KS 1731-260, RX J1709-2639, SAX 1810.8-2609
and 1H 1905+000 are taken from
\cite{tomsick04}, \cite{cackett06},
Jonker et al.\ (2003, 2004 and 2006), respectively.

Fig.\ \ref{Fig:exot} shows thermal states of nonsuperfluid
NSs with three different compositions in the inner core.
In both panels, the upper (basic) dotted curve
corresponds to low-mass NSs which possess no
inner core and cool slowly via the modified Urca
process from the outer core. The basic curve is almost
independent of NS mass $M$ as long as the inner core is
absent. For higher $M$, we obtain noticeably colder NSs,
cooling via enhanced neutrino emission from the
inner core. Their surface temperature strongly depends on the composition
in the inner core.
The coldest is the maximum-mass NS
(solid lines).

The coldest INSs observed to date
are consistent with all three neutrino emission scenarios
(see \citealt{ylh03}, for details of the models). At
present, the data on SXTs contain colder sources and
seem to be more restrictive.
From the upper limits on the thermal
luminosity of SAX 1808.4-3658 \citep{camp02} and 1H
1905$+$000 \citep{jonker06}, one can infer that a dominant process in
superdense matter should be more powerful
than direct-Urca-type processes with
kaons or pions. However, these results should be taken with caution --
some issues of theory and observations of SXTs still have to
be clarified (Sect.\ \ref{Sect:Disc}).

\begin{figure*}[th]
\begin{picture}(210,80)(30,30)
\put(29,-22){\includegraphics[width=17.5cm]{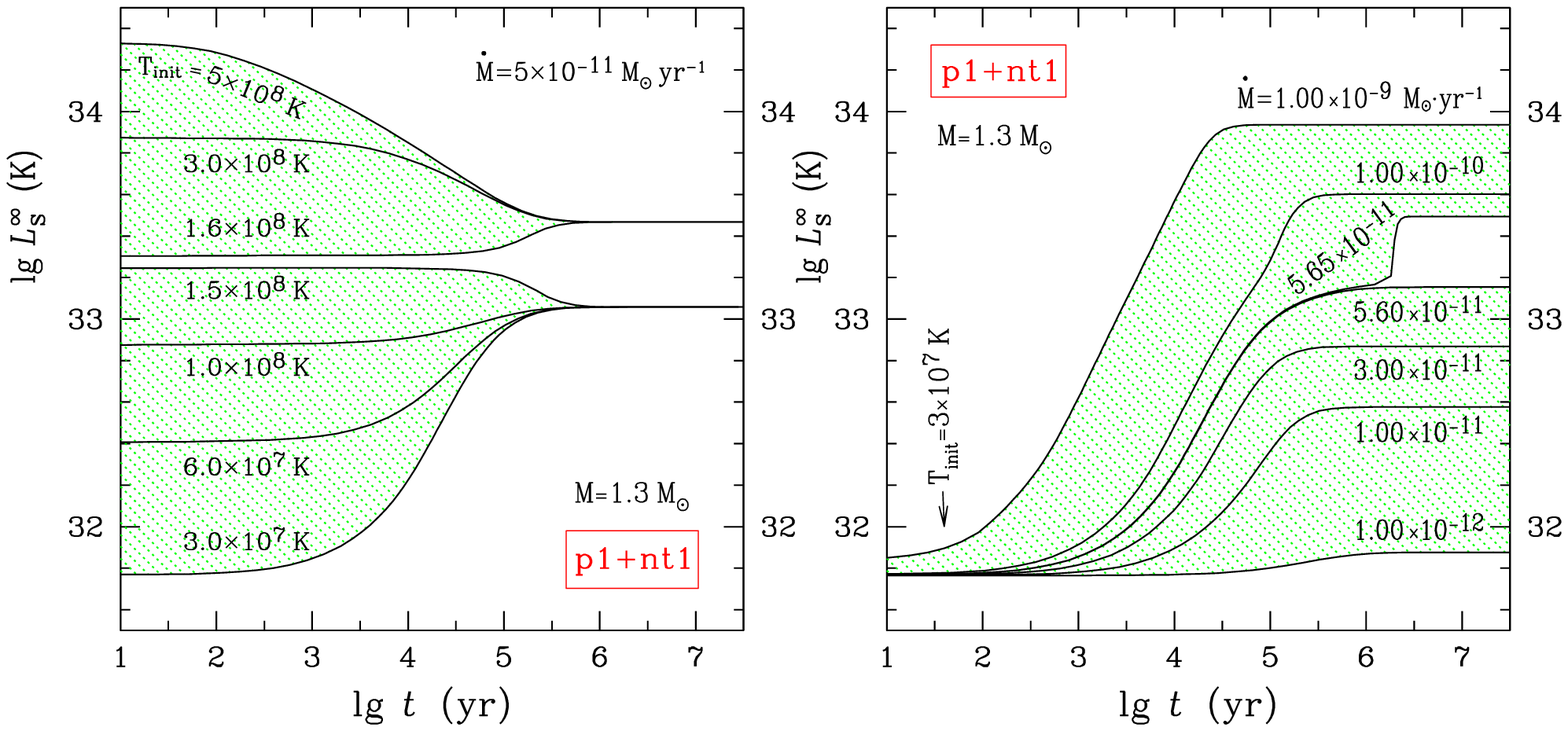}}
\end{picture}
\caption{\hspace{.2cm} Thermal evolution of NSs
($M=1.3\,M_\odot$) in
SXTs from an initial state to their final steady states.
Left: evolution, at a given
mass accretion rate, for several initial values
of the inner stellar temperature.
Right: evolution from a given initial state for several values of the
mean mass accretion rate. }
\label{Fig:split}
\end{figure*}

In what follows we will limit ourselves to the simplest nucleon
models of NS cores descibed, e.g., in Yakovlev \& Pethick (2004).
Results presented in
Figs.~\ref{Fig:none}--\ref{Fig:split} are obtained
using our generally relativistic code of NS thermal evolution.
We assume (for clarity of our study)
the absence of light-element accreted envelopes
on NS surfaces and neglect the effects of NS magnetic fields.
NS models in  Figs.~\ref{Fig:none}--\ref{Fig:1p1nt}
are based on a moderately stiff EOS
proposed by \cite{prakash88}.
This EOS opens the nucleon direct
Urca process at $\rho
\ge 2.8\rho_0$, i.e.,
in the inner cores of massive NSs
ith $M\geq M_D=1.358\, M_\odot$.

According to Fig.~\ref{Fig:none}, models of
non-superfluid NSs with enhanced emission in the inner cores
cannot explain the data on INSs and
SXTs. First, they are unable to interpret hottest sources.
Second, a transition between widely spaced
hot and cold NSs thermal states
occurs within an unrealistically narrow
NS mass range
$\sim 0.01\,M_\odot$. As we show below, including superfluidity
relieves these shortcomings.


\section{Superfluidity of nucleon matter}
\label{Sec: Superfluidity}

Microscopic theories of dense nucleon matter predict that
below some critical temperature, neutrons and protons in
NS cores are superfluid.
However, the critical temperatures $T_{cn}$ and $T_{cp}$ as
a function of density are very uncertain; see, e.g., \cite{ls01}.
Therefore, at present, it seems reasonable
to rely on a few general points of recent  theories
of  nucleon superfluidity:
\vskip 2mm
\begin{itemize}
\item proton pairing occurs in the $^1S_0$ state and persists
from  $\sim 0.5\rho_0$ up to a few $\rho_0$;

\item  critical temperature for protons, $T_{cp}(\rho)$, may be
rather high, with the maximum $\ga 10^9$~K somewhere between
$\rho_0$ and $2\rho_0$;

\item neutrons form  pairs in a $^3P_2$
state; this pairing is, typically, weaker and persists to higher
densities than the proton one;

\item critical temperature for
neutrons, $T_{cn}(\rho)$, has maximum somewhere  between $\rho_0$ and
few $\rho_0$; as a rule, this maximum  is shifted toward higher $\rho$
relatively to the maximum of $T_{cp}(\rho)$.

\end{itemize}
\vskip 2mm

These features of $T_{cp}(\rho)$ and $T_{cn}(\rho)$
can be simulated with phenomenological models.
We adopt the models shown in Fig.\ \ref{Fig:sf}.

\begin{figure*}[th]
\begin{picture}(210,80)(30,30)
\put(29,25){\includegraphics[width=17cm]{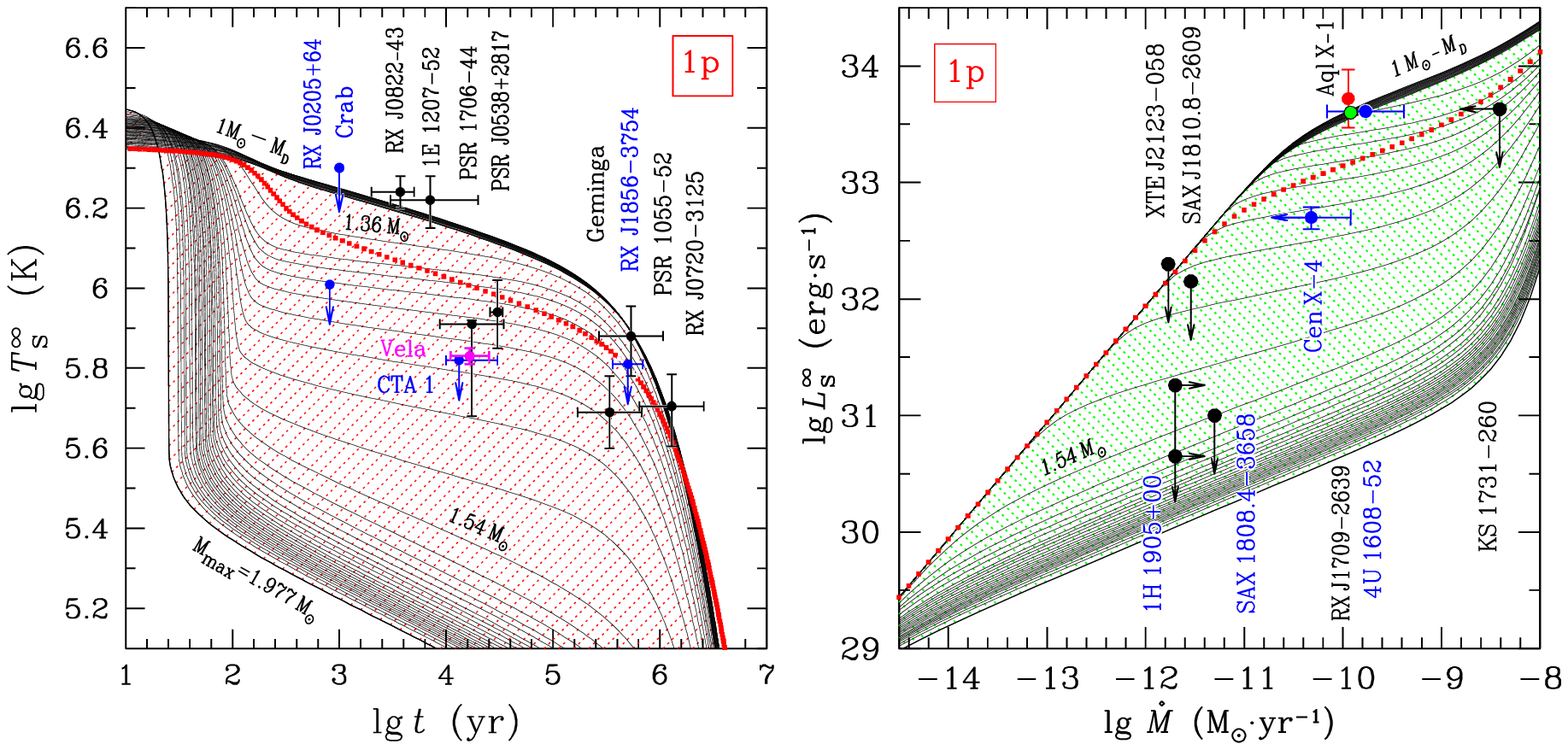}}
\end{picture}
\caption{Same as in Fig.~\protect{\ref{Fig:none}} but for
         strong proton superfluidity in the NS core (model ``1p'' 
	 in Fig.\ \ref{Fig:sf}).
	 In this case, $T_{cp}(\rho)$ has maximum $\approx 6.9\times 10^9$~K 
	 at $\rho \sim 2\,\rho_0$, remains $\sga 5\times 10^8\,$K at 
	 $\rho \sla 3.2\,\rho_0$, and dies out at $\rho \ssim 3.3\,\rho_0$.
         Accordingly, the entire cores of NSs with $M\sla 1.49\,M_\odot$ are
         strongly superfluid, while the inner cores of NSs with
         $M \sga 1.52\,M_\odot$ have nonsuperfluid central kernels.
} \label{Fig:1p}
\end{figure*}

\paragraph{\bf Strong proton superfluidity.}                
\label{Subsec:1p}

According to several authors (see \citealt{kyg02} for references)
a nucleon NS model with the open direct Urca in the inner core
and strong proton pairing in the outer core can explain available 
observations of INSs. Here we show that this model can explain also
the data on SXTs. Our results are displayed in Fig.~\ref{Fig:1p}.

The effect of strong proton superfluidity is twofold. First, this
superfluidity suppresses neutrino emission from NSs, making them hotter 
at a given age or mass accretion rate. This brings thermal states of
slowly cooling low-mass NSs  into the agreement with observations of
hotter sources. In massive NSs with enhanced cooling,
strong proton superfluidity may spread out the opening of
the direct Urca process over some density range.
As a  result, the enhanced coo\-l\-ing sets in gradually
with the growth of NS mass (not as sharp at $M=M_D$, as in 
Fig.~\ref{Fig:none}). This allows one to interpret the colder sources 
as massive NSs of different masses.

Strong proton superfluidity, with $T_{cp}\! \ga \!7\times 10^9\,$K,
appears in hot NSs with large neutrino luminosity, so
that the neutrino emission due to Cooper pairing of protons is
unimportant 
(Yakovlev et al. 2001).
However, this sup\-er\-flu\-i\-di\-ty suppresses the
powerful Urca processes. Now,  low-mass NSs mainly cool via a much weaker
neut\-ron\--neu\-tron brems\-strah\-lung (unaffected by proton
pairing). It slows down the cooling and let the low-mass stars be hotter
at a given $t$ or $\dot{M}$. In this way one can interpret observations 
of the hotter sources without invoking any reheating mechanism.
As noted by
Yakovlev et al. (2004),
the presence of light elements
(H, He) in the NS heat\--blan\-keting envelope facilitates
interpretation of the hotter INSs. Such an envelope is more 
heat transparent and let the NS look hotter for a given inner
temperature. Moreover, in that case even a weaker
superfluidity (with $T_{cp} \ga 10^9$~K) allows us to interpret 
the hotter sources; we have checked that this is also true for SXTs.

\begin{figure*}[t]
\begin{picture}(210,80)(30,30)
\put(26,15){\includegraphics[width=17cm]{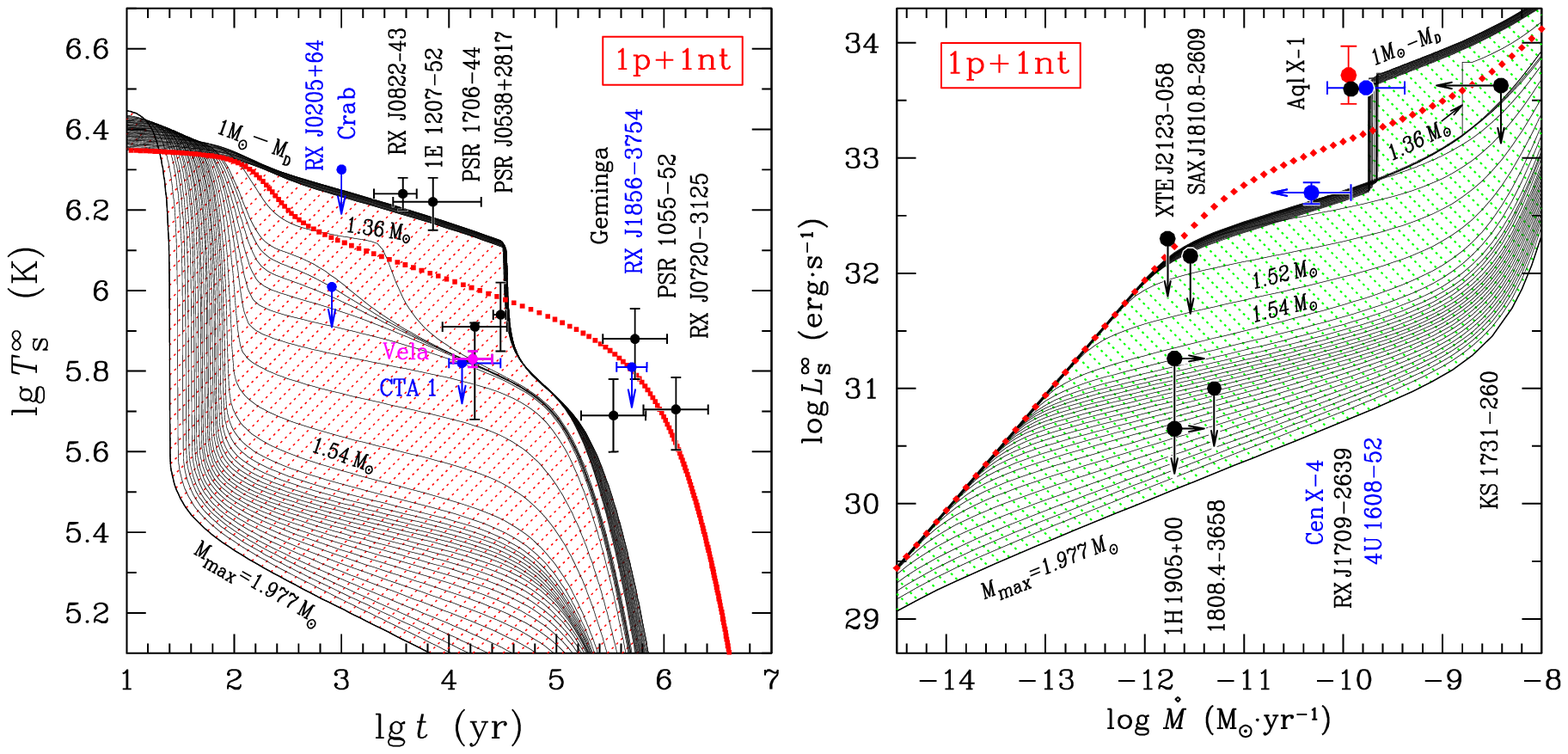}}
\end{picture}
\caption{\hspace{.2cm} Same as in Fig.~\protect{\ref{Fig:1p}}
   but with an addition of mild neutron superfluidity in the outer 
   NS core (model ``1nt'' 
   in Fig.\ \ref{Fig:sf}).
   In this case,  $T_{cn}(\rho)$  has maximum at $\rho\ssim 2\,\rho_0$ 
   and  dies out at $\ssim 5.7\,\rho_0$ 
   (i.e., neutrons are nonsuperfluid in the central kernels 
    of NSs with $M>1.89 \, M_\odot$).
   The maximum of the $T_{cn}(\rho)$ curve is wide --
   $T_{cn}(\rho)$ remains within $(2.0$--$3.3)\times 10^8\,$K at
   $\rho \ssim (0.72$--$3.2)\rho_0$. This density range
   corresponds to central densities of low-mass NSs
   ($M<1.358\,M_\odot$) and moderately  massive NSs (with
   $M\sapprox (1.36$--$1.48)\,M_\odot$). }
\label{Fig:1p1nt}
\end{figure*}

Thermal evolution of slowly cooling low-mass NSs is almost
independent of the assumed EOS (Page \& Applegate 1992),
as well of the NSs mass. At a given age or mass
accretion rate, all low-mass stars have nearly the same inner
temperature. Consequently, the appearance of
superfluidity affects all these stars in the same way;
cf.\ Figs.~\ref{Fig:none} and \ref{Fig:1p}
(and Fig.~\ref{Fig:1p1nt} below). This property holds as long as
proton pairing is strong in the entire NS core.

The situation is different in more massive NSs.
The impact of proton superfluidity on these stars depends also
on how far proton pairing extends into the inner core
and how steep is the slope of the $T_{cp}(\rho)$ profile
in this superfluid region
\citep{yh03}.
In Figs.~\ref{Fig:1p} and \ref{Fig:1p1nt}
the inner core of NS models with  $M_{\rm D} \le M \le M_{max}$
may occupy densities from $2.8\rho_0$ to $9.2\rho_0$
being superfluid at $\sim (2.8$--$3.3)\rho_0$ and normal at higher $\rho$.
Accordingly, in medium-mass NSs with $M \sim (1.36$--$1.52)\, M_\odot$
the inner core is entirely superfluid.
Superfluidity is mild near the outer core boundary,
and weakens rapidly toward the core center. The more massive
the star, the weaker proton superfluidity in its central part and the less
it suppresses the direct Urca process, thus letting the star
cool down faster. In this way thermal states of medium-mass INSs
become dependent of $M$ \citep{khy01}.

In the innermost central kernels of most massive NSs ($M\!>\! 1.52\, M_\odot$)
proton superfluidity dies
out, and the direct Urca process
reaches its full power.
This powerful fast cooling,  even in a small
part of the core,
renders the superfluid NS as cold as the nonsuperfluid one;
cf.\ Figs.~\ref{Fig:none} and \ref{Fig:1p}.

\paragraph{\bf Mild neutron superfluidity in the outer NS core. }
\label{Subsec: 1nt_outer}
Let us now  add mild neutron superfluidity and
assume first that this superfluidity
is located in the outer core.
We have adopted  model ``1nt'' shown in Fig.~\ref{Fig:sf}.
Cooling of INSs with superfluidities ``1nt'' and ``1p'' was
studied by \cite{khy01} and was shown to be
inconsistent with hotter sources.
We have tested this statement against the data on SXTs
with the same conclusion. Results are displayed in
Fig.~\ref{Fig:1p1nt}.

%
\begin{figure*}[th]
\begin{picture}(210,80)(30,30)
\put(28,15){\includegraphics[width=17cm]{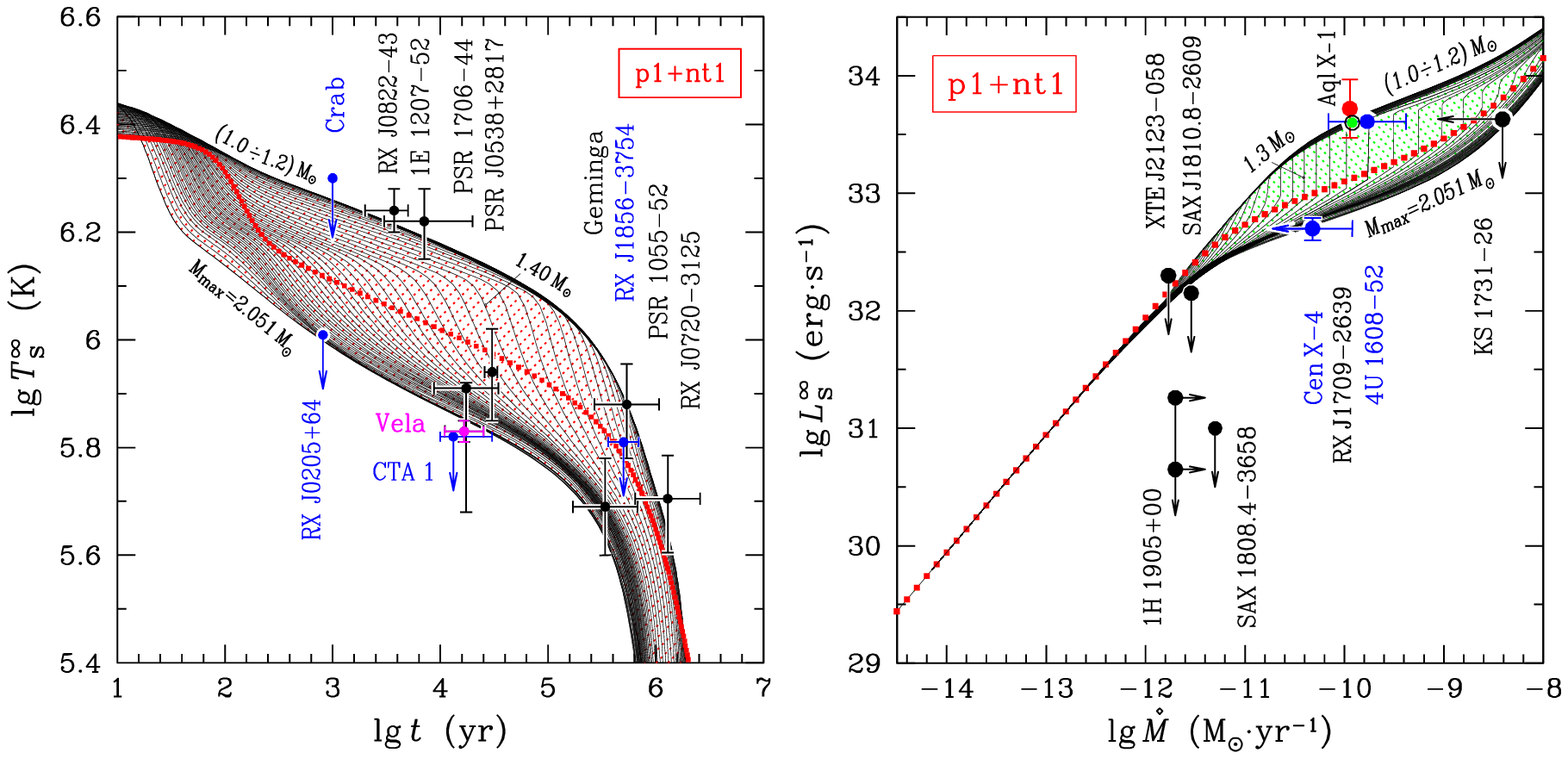}}
\end{picture}
\caption{\hspace{.2cm} 
   Thermal states of superfluid NSs based on the EOS of \cite{dh01} 
   which prohibits the direct Urca process in NS cores. 
   The figure illustrates the so called "minimal cooling scenario" 
   in which  Cooper pairing of neutrons operates as enhanced cooling agent. 
   The scenario assumes
   strong proton pairing in the outer NS core and mild neutron pairing 
   in the inner core (models ``p1'' and ``nt1'' 
   in Fig.\ \ref{Fig:sf}, respectively).
   In both panels the dotted basic curve refers  to the same 
   cooling scenario as in Fig.~\protect{\ref{Fig:exot}}.
   From top to botton, thin solid  curves show NSs models with 
   masses growing from $1\,M_\odot$ to $M_{max}=2.051\,M_\odot$,  
   with a step of $0.02\,M_\odot$.
	Pairing ``p1''  has a wide maximum at $\rho \ssim 1.6\rho_0$ 
   and dies out at $\ssim 3.8\,\rho_0$, i.e.\ in the central kernels of 
   NSs with $M \sga  1.49\,M_\odot$.
   Strong pairing with $T_{cp} \sapprox(2.0$--$6.9)\times
   10^9\,$K persists up to  $\rho \ssim 3.2\,\rho_0$ and
  extends over entire cores of NSs with $M \sla 1.27\,M_\odot$.
      Neutron pairing of ``nt1'' is weak in the outer core, at 
  $\rho \sla 3.1\,\rho_0$ (i.e., in NSs with $M\sla 1.24 M_\odot$), 
  has a sharp maximum of mild strength in the inner core, at 
  $\rho \ssim 4.7\, \rho_0$, and dies out rapidly as $\rho$ approaches 
  $\ssim 6.7\,\rho_0$.
  Mild pairing with $T_{cn}\sapprox (2.0$--$6.0)\!\times \! 10^8\,$K extends 
  over the density interval $\ssim (3.1$--$6.4)\rho_0$, which corresponds to 
  the inner cores of NSs with $M\sapprox  (1.22$--$1.95)M_\odot$ or to
  spherical layers around the nonsuperfluid central kernels of
  NSs with $M\sapprox (1.97$--$2.05)M_\odot$. 
}
\label{Fig:min}
\end{figure*}
%

In cooling INSs the effect of mild neutron superfluidity
is very spectacular. First, neutron pairing reduces the heat
capacity of NSs, because the heat is stored mostly in neutrons.
The reduction amounts to a factor of several in low-mass NSs.
This makes low-mass 
 NSs
very cold after appearing of such pairing
 at $t\ga 3\times 10^4$~yr.
The cooling  is additionally accelerated by powerful neutrino
emission, triggered by Cooper pairing of neutrons.
In result, the low-mass NSs models become unable to interpret four
old sources: Geminga, RX J1856--3754, PSR 1055--52 and RX~J0720--3125.

In the most massive NSs with $M>1.54 \, M_\odot$ 
the effect of neutron superfluidity is opposite.
In these stars the mild pairing appears earlier, at $t\sim 10^2$ yr,
when neutrino emission due to the direct Urca process is
much stronger than the Cooper pairing neutrino emission (which is,
therefore, insignificant).
On the other hand, additional suppression of the
direct Urca process by neutron superfluidity
makes the most massive NSs a bit hotter in the advanced stage of neutrino
cooling era.

As for transiently accreting NSs in SXTs, their ther\-mal states
become independent of the NS heat capacity as soon as  they
reach the steady-state regime (Yakovlev et al.\ 2003; 2004b).
Hence, a strong reduction of the heat capacity does not affect
NS thermal states at the photon-dominated stage. That is why  the slope
of the dot\-ted curve at
$\dot{M}\!\la\! 10^{-12}\,M_\odot\cdot\mbox{yr}^{-1}$,
corresponding to this stage, remains unchanged. However,
at the neutrino stage,
neutrino emission due to Cooper pairing of neutrons makes
low-mass SXTs much colder (which unables one to
interpret such hot sources as Aql X-1 and RX 1709--2639).
The same effect in
lightest among the massive SXTs hampers the explanation of  KS
1721--260. Similarly to high-mass INSs, high-mass SXTs become a
bit hotter;  neutron superfluidity, in its turn,
slightly broadens the span of their allowed thermal states,
stretched before by the proton pairing.

\paragraph{\bf Mild neutron superfluidity in the inner NS core.}
\label{Subsec: 1nt_inner}

The presence of mild neutron superfluidity in the outer NS core
contradicts observations of INSs and SXTs, as it renders
low-mass NSs too cold. However,
mild pairing might occur in the inner core which is present
only in massive NSs. Let us explore this case.

From Fig.~\ref{Fig:1p1nt} we see that neutrino
emission due to Cooper pairing of neutrons
may initiate a very fast cooling and, thus, may serve
as a fast cooling agent (at least, for INSs).

In INSs, such a possibility was studied by \cite{plps04}
and \cite{gkyg04}. In particular, the latter authors
consider nucleon NS models based on the EOS of \cite{dh01}
which prohibits the direct Urca process. Their cooling scenario
is based on strong proton pairing in the outer NS core and mild
neutron pairing in the inner core (phenomenological models
``p1'' and ``nt1'', respectively;
see Fig.\ \ref{Fig:sf}). 
The model ``nt1'' has a specific dependence $T_{cn}(\rho)$
which keeps neutron pairing weak in the outer core. Let us
remind that weak superfluidity, with $T_{cn} \la 2\times 10^8\,$K,
does not appear in low-mass NSs in the neutrino cooling era:
these NSs are too hot. Accordingly, in this era neutron pairing
``nt1'' affects only massive NSs.

\cite{gkyg04} compared such cooling scenario
(called by Dany Page a ``minimal cooling
model", for its simplicity, and the minimal number of its
ingredients) with observations of INSs, and found it marginally
consistent with them. The scenario can explain those
INSs whose thermal emission is detected with confidence,
as illustrated in the left panel of Fig.~\ref{Fig:min}.
However, these authors  noted that the model will fail if
the two sources, RX J0205+64 and CTA$\,1$ (expected
to be thermally emitting INSs), turn out to be much colder
than their presently established upper limits, or if very
cold INSs are  detected in future.

Since NSs in SXTs are expected to be rather massive, owing to
accretion of matter from their  companions, it is worth to test the
``minimal cooling scenario" against the data on SXTs. That is done in
the right panel of Fig.~\ref{Fig:min}. One can see that the
model successfully reproduces hotter SXTs,
including the frequently bursting sources Aql$\,$X-1, RX 1709--2639 and
4U 1608--52. However, the upper limits for at least two sources,
1H 1905+000 and SAX 1808.4--3658, definitely
fall far below  the predictions of the model.

The transient source 1H$\,$1905+000  was recently observed with Chandra
\citep{jonker06}; the upper limit for its quiescent thermal luminosity
seems to be firm. On the contrary, the value of $\dot{M}$
for this object is quite uncertain; some physical
arguments allow one to believe it to be higher than
$10^{-12}\,M_\odot\cdot\mbox{yr}^{-1}$. For the frequently bursting
transient SAX 1808.4--3658 the mass accretion rate is known more accurately,
while the upper limit of its quiescent thermal luminosity
is less certain.
It varies from $\sim 6\times 10^{29}$ erg$\cdot$s$^{-1}$ in \cite{camp02}
to $\sim 4\times 10^{31}$ erg$\cdot$s$^{-1}$ in estimations of P.\ Shtykovski
(private communication; see details in \citealt{yak04b} and \citealt{yp04}).
Never\-the\-less, both these upper limits
are clearly much below the predictions of the minimal cooling model.
In present paper we have adopted the same upper limit
on  bolometric luminosity of SAX 1808.4--3658
as in \cite{ylh03}.

Thus, despite many theoretical and observational uncertainties, it
seems that the ``minimal cooling model" is ruled out by the data on
SXTs. Therefore, the mechanism of the
enhanced NS cooling should be more powerful than the
process of Cooper pairing of neutrons.

Nevertheless, mild neutron superfluidity localized in the inner
NS core is not prohibited. If it is available, together
with proton superfluidity,
the enhanced cooling should be of the direct Urca type.
If, however, proton pairing does not extend to the inner
core, neutron superfluidity remains the  only regulator of
cooling of massive NSs. This case was studied by \cite{gkyg05}
in NS models with open direct Urca process.
Because of the lack of space, we do not illustrate this case. We just note
that this model easily  fits hot and cold sources but
scarcely explains quite a large span of
intermediate sources, for example,  Vela and Cen~X-4.

\section{Dichotomy of thermal evolution of SXTs?}

As mild neutron superfluidity might exist in the inner NS core,
let us outline its effect on thermal evolution of
SXTs. In contrast to INSs, thermal states of NSs in SXTs
depend not only on NS properties but also on the accretion rate.

As seen from Fig.~\ref{Fig:min} 
(right), heating curves for medium-mass
SXTs
drop abruptly when the mass accretion rate becomes lower
than some threshold value specific for an NS of a given mass. As we already
know, this drop 
(the vertical segment of a curve) 
signals the onset of neutron superfluidity
in the NS core when
the deep crustal heating becomes insufficient to keep the core
temperature above the maximal critical temperature of neutrons.

Let us remind that NSs are thermally inertial objects. A thermal
state of the star as a whole may change noticeably on time
scales of $\sim 10^4$ yr \cite{colpi01}. We also recall that
heating curves refer to steady-states which NSs
reach in a few
millions years after the onset of the accretion stage in
SXTs. The abrupt drops of heating curves indicate that
thermal evolution of SXTs exhibits {\it dichotomy}. This means that
two NSs with the same mass may evolve to very different steady states
if their mass accretion rates are slightly different.
The right panel of Fig.~\ref{Fig:split} illustrates how
a change of $\dot{M}$ by  one percent from
some ``threshold" value entails a change of the
steady state thermal luminosity by a factor of $\sim 1.5$.

Moreover, a tiny mismatch of masses of two NSs, or a
small difference of their inner temperatures before the onset of the
accretion stage, may also induce
similar dichotomy. The latter effect is illustrated in
the left panel of Fig.~\ref{Fig:split}. The former one can be
seen in Fig.~\ref{Fig:min} (and Fig.~\ref{Fig:1p1nt}):
at some "threshold" $\dot{M}$,  NSs with
masses different by one percent
($\sim 0.02\,M_\odot$) may show thermal luminosities which differ
by factor of $\sim 3$. One can also speculate that the dichotomy can be
caused by a variable crust composition,
which can change the efficiency
of deep crustal heating at a given $\dot{M}$
(see Haensel \& Zdunik, 1990; 2003). Concluding, whatever causes
the dichotomy, the presence of mild neutron superfluidity
in the NS core may considerably complicate an interpretation of thermal
emission from SXTs.

\section{Discussion}
\label{Sect:Disc}

We have shown that the data on quiescent thermal emission from SXTs require
the presence of strong proton superfluidity and the absence of
mild neutron superfluidity in the outer NS core (i.e., in
low-mass NSs). The mechanism which controls fast neutrino cooling of
massive NSs should be more powerful than that due to Cooper pairing of
neutrons and, probably, than the direct-Urca-type processes with kaons
and pions. We have noticed that the presence  of mild neutron
superfluidity in the inner NS core may cause the dichotomy of
thermal evolution of SXTs.

Further studies are required to confirm or reject our inferences. The
data on SXTs are very uncertain. Much work is needed
to constrain these data and clarify the nature of variability
of some SXTs in quiescence. There are still many challenges
to respond to in order to build a realistic theory of deep
crustal heating of SXTs.
Note that current models of accreted crust and deep
crustal heating are based on one particular cold liquid drop model
of atomic nuclei and one model of pycnonuclear burning. The actual
properties of highly neutron-rich nuclei surrounded by free neutrons
in the inner NS crust, and the rate of pycnonuclear fusion
of such nuclei (Coulomb barrier penetrability, astrophysical
$S$-factors) are known with considerable uncertainty
(see, e.g., \citealt{ywg06}).
In this study we have neglected
neutrino emission due to Cooper pairing of neutrons in the NS crust
in order to show  directly the effect of proton and neutron superfluidity
in the NSs core on thermal states of SXTs.

\begin{figure}[th]
\begin{picture}(110,65)(30,30)
\put(17,1){\includegraphics[width=9.5cm]{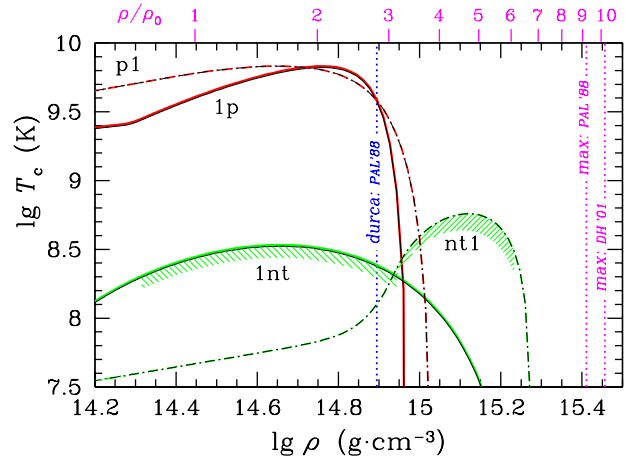}}
\end{picture}
\caption{\hspace{.2cm} Phenomenological models of neutron and proton 
superfluidities. {\sl Solid curves}: models  "1nt"  and "1p" from \cite{khy01}, 
used in the cooling scenarios with the direct Urca process
in Figs.\ \ref{Fig:1p} and \ref{Fig:1p1nt}.
{\sl Dot-dashed curves}: models "nt1" and  "p1" from \cite{gkyg04},
adopted in the "minimal cooling" scenario in Fig.\ \ref{Fig:min}.
{\sl Vertical dotted lines}:
the maximum-mass NS central densities for EOSs 
by Prakash et al.\ (1988) and by Douchin and Haensel (2001), 
used in the former and the latter scenarios, respectively,
as well as the direct Urca threshold (for {\it PAL'88}).
}
\label{Fig:sf}
\end{figure} 

\begin{acknowledgement}
\hspace*{.5cm}\\
 We express our gratitude to  D.G.Yakovlev for sharing with us
 his expertise,  and for his help in the preparation of this
 paper. We thank  A.D.Kaminker and
 M.E.Gusakov
  for providing us with the table
of EOS without dUrca. KL is grateful
the organizers of the conference ``Isolated Neutron Stars: from
the Interior to the Surface'' for financial support.
\end{acknowledgement}


\begin{thebibliography}{3}

\bibitem[\protect\citeauthoryear{Brown et al.}{1998}]{brown98}
 Brown E.F., Bildstein L., Rutledge R.E. ApJ, {\bf 504}, L95  (1998)

\bibitem[\protect\citeauthoryear{Cackett et al.}{2006}]{cackett06}
Cackett E.M., Wijnands R., Linares M. et al. MNRAS, {\bf 369}, 407 (2006)

\bibitem[\protect\citeauthoryear{Campana et al.}{2002}]{camp02}
Campana S., Stella L., Gastadello F. et al. ApJ., {\bf 575}, L15 (2002)


\bibitem[\protect\citeauthoryear{Douchin \& Haensel}{2001}]{dh01}
    Douchin F.\ \& Haensel P. A\&A, {\bf 380}, 151 (2001)

\bibitem[\protect\citeauthoryear{Colpi et al.}{2001}]{colpi01}
  Colpi M., Geppert U., Page D. et al. ApJ, {\bf 548}, L175 (2001)


\bibitem[\protect\citeauthoryear{Gusakov et al.}{2004}]{gkyg04} 
 Gusakov M.E.,  Kaminker A.D., Yakovlev D.G. et al.
 A\&A, {\bf 423}, 1063 (2004)


\bibitem[\protect\citeauthoryear{Gusakov et al.}{2005}]{gkyg05}
 Gusakov M.E., Kaminker A.D., Yakovlev D.G., Gnedin O.Y.
 MNRAS, {\bf 363}, 555 (2005)

\bibitem[\protect\citeauthoryear{Haensel \& Zdunik}{1990}]{hz90}
Haensel P.\ \& Zdunik J. L.
A\&A, {\bf 227}, 117 (1990)

\bibitem[\protect\citeauthoryear{Haensel \& Zdunik}{1990}]{hz03}
Haensel P.\ \& Zdunik J. L.
A\&A, {\bf 404}, L33 (2003)


\bibitem[\protect\citeauthoryear{Jonker et al.}{2003}]{jonker03}
Jonker P.G., Mendez M., Nelemans G. et al. MNRAS, {\bf 341}, 823 (2003)


\bibitem[\protect\citeauthoryear{Jonker et al.}{2004}]{jonker04}
Jonker P.G., Wijnands R. \& van der Klis M. MNRAS, {\bf 349}, 94 (2004)

\bibitem[\protect\citeauthoryear{Jonker et al.}{2006}]{jonker06} 
  Jonker P.G., Bassa C.G., Nelemans G. et al.
  MNRAS, {\bf 368} 1803 (2006)


\bibitem[\protect\citeauthoryear{Kaminker et al.}{2001}]{khy01}      
 Kaminker A.D., Haensel P., Yakovlev D.G.    
 A\&A, {\bf 373}, L17 (2001)


\bibitem[\protect\citeauthoryear{Kaminker et al.}{2002}]{kyg02}  
  Kaminker A.D., Yakovlev D.G.\  \& Gnedin O.Y.
  A\&A, {\bf 383}, 1076 (2002)

\bibitem[\protect\citeauthoryear{Lombardo \& Schulze}{2001}]{ls01}
  Lombardo U.\ \& Schulze H.-J.
  In: Physics of Neutron Stars Interiors, eds.\ Blaschke D..
     Glendenning N.K., Sedrakian A., pp.\ 30--53. Berlin,
     Springer-Verlag, 2001


\bibitem[\protect\citeauthoryear{Page \& Applegate}{1992}]{pa92}
 Page D.\ \& Applegate J.H. ApJ, {\bf 394}, L17 (1992)

\bibitem[\protect\citeauthoryear{Page et al.}{2004}]{plps04}
 Page D., Lattimer J.M.\ \& Prakash M. et al.
 ApJ Suppl., {\bf 155}, 623 (2004)     

\bibitem[\protect\citeauthoryear{Page et al.}{2006}]{pgw06}   
  Page D., Geppert U. \& Weber F., Nucl.\ Phys.\ A, {\bf 777}, 497 (2006) 

\bibitem[\protect\citeauthoryear{Prakash et al.}{1988}]{prakash88}
Prakash M., Ainsworth T.L. \& Lattimer J.M. Phys.\ Rev.\ Lett.,
   {\bf 61}, 2518 (1998)

\bibitem[\protect\citeauthoryear{Tomsick et al.}{2004}]{tomsick04}
Tomsick J.A., Gelino D.M., Halpern J.P. et al. ApJ, {\bf 610}, 933 (2004)


\bibitem[\protect\citeauthoryear{Yakovlev et al.}{2001}]{ykgh01}
  Yakovlev D.G., Kaminker A.D.,  Gnedin O.Y.\  \&  Haensel P.
  Phys.\ Rep., {\bf 354}, 1 (2001)                               

\bibitem[\protect\citeauthoryear{Yakovlev \& Haensel}{2003}]{yh03}
  Yakovlev D.G.\  \&  Haensel P. A\&A, {\bf 407}, 259 (2003)    

\bibitem[\protect\citeauthoryear{Yakovlev \& Pethick}{2004}]{yp04}
 Yakovlev D.G., Pethick C.J. Ann.\ Rev.\ Astron.\ Astrophys.,
 {\bf 42}, 169 (2004)

\bibitem[\protect\citeauthoryear{Yakovlev et al.}{2003}]{ylh03}
   Yakovlev D.G., Levenfish K.P. \& Haensel P. A\&A, {\bf 407}, 265 (2003)

\bibitem[\protect\citeauthoryear{Yakovlev et al.}{2004}]{yak04b}
   Yakovlev D.G., Levenfish K.P., Potekhin A.Y. et al. A\&A,
   {\bf 417}, 169 (2004)

\bibitem[\protect\citeauthoryear{Yakovlev et al.}{2006}]{ywg06}
   Yakovlev D.G., Weischer M. \& Gasques L. MNRAS, {\bf 371}, 1322 (2006) 

\end{thebibliography}
\end{document}